\documentclass[aps, showpacs,preprintnumbers, superscriptaddress, nofootinbibt, twocolumn]{revtex4-2}
\usepackage{eurosym}
\usepackage{graphicx}
\usepackage{amssymb}
\usepackage{amsmath}
\usepackage{color}
\def\be{\begin{equation}}
\def\ee{\end{equation}}
\def\bea{\begin{eqnarray}}
\def\eea{\end{eqnarray}}

\begin{document}

\title{The first variation of the matter energy-momentum tensor with respect to the metric, and its implications on modified gravity theories}
\author{Zahra Haghani}
\email{z.haghani@du.ac.ir}
\affiliation{School of Physics, Damghan University, Damghan, 41167-36716, Iran}
\author{Tiberiu Harko}
\email{tiberiu.harko@aira.astro.ro}
\affiliation{Department of Physics, Babes-Bolyai University, 1 Kogalniceanu Street,
	400084 Cluj-Napoca, Romania,}
\affiliation{Department of Theoretical Physics, National Institute of Physics
	and Nuclear Engineering (IFIN-HH), Bucharest, 077125 Romania,}
\affiliation{Astronomical Observatory, 19 Ciresilor Street, 400487
	Cluj-Napoca, Romania,}

%\affiliation{School of Physics, Sun Yat-Sen University,  510275 Guangzhou, People's
	%Republic of China.}
\author{Shahab Shahidi}
\email{s.shahidi@du.ac.ir}
\affiliation{School of Physics, Damghan University, Damghan, 41167-36716, Iran}

\begin{abstract}
The first order variation of the matter energy-momentum tensor $T_{\mu \nu}$ with respect to the metric tensor $g^{\alpha \beta}$ plays an important role in modified gravity theories with geometry-matter coupling, and in particular in the $f(R,T)$ modified gravity theory. We obtain the expression of the variation $\delta T_{\mu \nu}/\delta g^{\alpha \beta}$ for the baryonic matter described by an equation given in a parametric form, with the basic thermodynamic variables represented by the particle number density, and by the specific entropy, respectively. The first variation of the matter energy-momentum tensor turns out to be independent on the matter Lagrangian, and can be expressed in terms of the pressure, the energy-momentum tensor itself, and the matter fluid four-velocity. We apply the obtained results for the case of the $f(R,T)$ gravity theory, where $R$ is the Ricci scalar, and $T$ is the trace of the matter energy-momentum tensor,  which thus becomes a unique theory, also independent on the choice of the matter Lagrangian. A simple cosmological model, in which the Hilbert-Einstein Lagrangian is generalized through the addition of a term proportional to $T^n$ is considered in detail, and it is shown that it gives a very good description of the observational values of the Hubble parameter up to a redshift of $z\approx 2.5$.
\end{abstract}

\date{\today}
\maketitle

%\tableofcontents

\section{Introduction}

There are at least three theoretical perspectives \cite{0} that could be used to explain the large amount of recent observations, which strongly suggest a faster and faster expanding Universe \cite{1,1a}, with a composition in which ordinary matter represents only 5\% of its composition, the rest being represented by the dark energy, and the dark matter \cite{1a,2}.
The first point of view is represented by the dark constituents theory, which adds two more components to the total energy momentum tensor of the Universe, representing dark matter and dark energy, respectively. Therefore the cosmological dynamics is described by the field equation $G_{\mu \nu}=\kappa ^2 T^{\rm bar}_{\mu \nu}+\kappa ^2T^{\rm DM}_{\mu \nu}(\phi, \psi _{\mu},...)+\kappa ^2T^{\rm DE}_{\mu \nu}(\phi, \psi _{\mu},...)$, where $T^{\rm bar}_{\mu \nu}$, $T^{\rm DM}_{\mu \nu}(\phi, \psi_{\mu},...)$, and $T^{\rm DE}_{\mu \nu}(\phi, \psi_{\mu},...)$ represent the energy-momentum tensors of baryonic matter, dark matter, and dark energy, respectively, with $\phi$ and $\psi_{\mu}$ representing scalar or vector fields. A well studied dark constituent model is represented by the quintessence (scalar field) description of dark energy \cite{Qu1,Qu2}.

In the dark geometry approach, an exclusively geometric attitude on the gravitational phenomena is adopted, by explaining the cosmological dynamics through the modification of the geometry underlying the Einstein field equations. Hence, the extended Einstein equations become in this approach  $G_{\mu \nu}=\kappa ^2T_{\mu \nu}^{{\rm bar}}+\kappa ^2 T_{\mu \nu}^{(\rm geom)}\left(g_{\mu \nu}, R, \square R,...\right)$, where $T_{\mu \nu}$ is the energy-momentum tensor of ordinary matter, and $T_{\mu \nu}^{(\rm geom)}\left(g_{\mu \nu}, R, \square R,...\right)$ is a purely geometric term, obtained from the metric, torsion $\tau$, nonmetricity $Q$, extensions of Riemann geometry etc., and which can effectively mimic dark energy, dark matter, or both. Some typical example of dark geometric theories are the $f(R)$ \cite{fR}, $f(Q)$ \cite{fQ}, hybrid metric-Palatini gravity \cite{HMP} theories, or gravitational theories based on the Weyl-Cartan-Weitzenb\"{o}ck \cite{Weyl} and Finsler geometries \cite{Fin1, Fin2}.

The third avenue for the understanding of the gravitational and cosmological phenomena is represented by the dark coupling approach, in which the standard Einstein gravitational equations are generalized to take the mathematical form  $G_{\mu \nu}=\kappa ^2T_{\mu \nu}+\kappa ^2 T_{\mu \nu}^{(\rm coup)}\left(R, L_m, T, \square R, \square T,... \right)$,  where the effective energy-momentum tensor $T_{\mu \nu}^{(\rm coup)}\left(g_{\mu \nu}, R, L_m, T, \square R, \square T,... \right)$ of the theory is built up by considering the maximal extension of the Hilbert-Einstein Lagrangian, by abandoning its additive structure in matter and geometry. In the dark coupling approach, matter is represented either by the trace $T$ of the matter energy-momentum tensor, by the matter Lagrangian $L_m$ or by some scalar made by $T_{\mu\nu}$ such as $T_{\mu\nu}T^{\mu\nu}$.

The dark coupling approach is also a theoretical answer to the problem of the maximal extension of the additive Hilbert-Einstein Lagrangian, which automatically implies a non-additive structure of the action in the geometric and matter variables.  In a general form the requirement of the maximal extension of the gravitational action can be implemented by assuming that the Lagrangian of the gravitational field is an arbitrary function of the curvature scalar $R$, and of the matter Lagrangian $L_m$.

One of the interesting features of the dark coupling models is that they imply the presence of a nonminimal geometry-matter coupling. Dark couplings are not restricted to Riemannian geometry, but they can be considered in the framework of the extensions of Riemann geometry. Typical examples of dark coupling theories are the $f\left(R,L_m\right)$ \cite{fRLM1,fRLM2}, $f(R,T)$ \cite{fRT}, $f\left(R,T, R_{\mu \nu}T^{\mu \nu}\right)$ \cite{fRTmunu1, fRTmunu2}, $f(\tau,T)$ \cite{fTT},  $f(Q,T)$ \cite{fQT}, $f\left(R,T,Q,T_m\right)$ \cite{fRTQ} and the teleparallel theories \cite{tele1, tele2,tele3,tele4}, respectively. For reviews on modified theories of gravity and dark energy see \cite{Od1, Cap,Od2, Lang, Fru, Bat}. Recently, conformally invariant gravitational theories have been also intensively investigated \cite{Man,Gh1,Gh2,Gh3}. Extensive studies of the Einstein-Gauss-Bonnet gravity were performed in \cite{G1,G2,G3}.

One of the interesting consequences of the dark coupling theories is the reconsideration of the role of the ordinary (baryonic) matter in the cosmological dynamics. Through its coupling to gravity, matter becomes a key element in the explanation of cosmic dynamics, and recovers its central role gravity, which is minimized or even neglected in the dark constituents and dark geometric type theories. An important implication of the geometry-matter coupling is that the matter energy-momentum tensor is generally not conserved, and thus an extra-force is generated, acting on massive particles moving in a gravitational field, with the particles following non-geodesic paths \cite{fRLM2,fRT}. The possibility of the existence of such couplings between matter and geometry  have opened interesting, and novel pathways  for the study of gravitational phenomena \cite{6}. The astrophysical and cosmological implications of the $f(R,T)$ gravity theory were investigated in \cite{F1, F2,F3,F4,F5,F6,F7,F8,F9,F10,F11}.

However, the dependence of the gravitational action in the dark coupling theories on $L_m$ gives a new relevance to the old problem of the degeneracy of the matter Lagrangian. Two, physically inequivalent expressions of the matter Lagrangian, $L_m=-\rho$, and $L_m=P$, lead to the same energy-momentum tensor for matter. This result has important implications for dark coupling gravity models. For example, in the framework of the $f\left(R,L_m\right)$ theory, it was shown in \cite{SF} that adopting for the Lagrangian density the expression $L_m=p$,  where $p$ is the pressure, in the case of dust the extra force vanishes. However, for the form $L_m=\rho$ of the matter Lagrangian,  the extra-force does not vanish \cite{BLP}. In \cite{H1} it was shown, by using the variational formulation for the derivation of the equations of motion, that both the matter Lagrangian, and the energy-momentum tensor, are uniquely and completely determined by the form of the geometry-matter coupling.  Therefore, the
extra-force never vanishes as a consequence of the thermodynamic properties of the system. In \cite{5} it was shown that if the particle number is conserved, the Lagrangian of a barotropic perfect fluid with $P=P(\rho)$  is $L_m=-\rho \left[c^2+\int{P(\rho)/\rho ^2d\rho}\right]$, where $\rho $ is the \textit{rest mass} density. This result can be used successfully in the study of the modified theories of gravity. The result is based on the assumption that the Lagrangian does not depend on the derivatives of the metric, and that the particle number of the fluid is a conserved quantity, $\nabla _{\mu}\left(\rho u^\mu\right)=0$.
The matter Lagrangian also plays an important role in the $f(R,T)$ theory of gravity \cite{fRT}.

In theories with geometry-matter coupling another important quantity, the variation of the energy-momentum tensor with respect to the metric does appear, and plays an important role. These variations  contribute with some new terms in the gravitational field equations. The corresponding second order tensor is denoted as $\mathbb{T}_{\mu \nu}$, and it is introduced via the definition   \cite{fRT}
$$ \mathbb{T}_{\mu\nu}\equiv g^{\rho\sigma}\frac{\delta T_{\rho\sigma}}{\delta g^{\mu\nu}}.$$
 If the matter Lagrangian does not depend on the derivatives of the metric, one can obtain for $\mathbb{T} _{\mu \nu}$ a mathematical expression that also contains the second variation of the matter Lagrangian with respect to the metric, $\delta ^2L_m/\delta g^{\mu \nu}\delta g^{\alpha \beta}$. The Lagrangian of the electromagnetic field is quadratic in the components of the metric tensor, and hence its second variation gives a non-zero contribution to $\mathbb{T}_{\mu \nu}$.

 However, the case of ordinary baryonic matter is more complicated. At first sight, by taking into account the explicit forms of the matter Lagrangians, $L_m=-\rho$, or $L_m=p$, no explicit dependence on the metric does appear, as opposed, for example,  to the case of the electromagnetic field. This would suggest that the second variation of the matter Lagrangian always identically vanishes, no matter what its functional form  is. This conclusion may be valid indeed for some special forms of the equation of state, but it is not correct if one adopts a general thermodynamic description of the baryonic fluids.

 The presence in the gravitational field equations of the terms related to the variation of the matter Lagrangian, and of the matter energy-momentum tensor, immediately lead to the important problem of the implications of the degeneracy of the matter Lagrangian in modified gravity theories. Interestingly enough, (at least) two distinct classes of models can be constructed for any modified gravity theory, based on the different choices of $L_m$. Even that many cosmological and astrophysical problems can be solved effectively within the framework of modified gravity theories by using this degeneracy, their theoretical basis, and understanding,  remain uncertain.

It is the goal of the present paper to investigate the problem of the second variation of the perfect fluid matter Lagrangian with respect to the metric tensor components, and to analyze its impact on modified gravity theories. As a first step in our analysis, we obtain, from general thermodynamic considerations, the expressions of the variations with respect to the metric  and of the baryonic matter energy density and pressure. Once these expressions are known, a straightforward calculation, involving the computation of the second variation of the energy density and pressure, gives the first variation of the matter energy-momentum tensor with respect to the metric, which also allows to obtain the tensor $\mathbb{T}_{\mu \nu}$.

The basic result of our investigation is that the tensor $\mathbb{T} _{\mu \nu}$ is {\it independent of the choice of the matter Lagrangian}. The effect of the second order correction is estimated in a cosmological background.

 In this context, the full clarification of the problem of the variation of the matter Lagrangian, and of the matter energy-momentum tensor, especially in modified gravity theories, is an important issue, whose solution could give new insights not only in the structure of modified gravity theories, but also open some new windows for the cosmological applications. In particular, the main novel aspect of our results is the proof of the uniqueness of the variation of the energy momentum tensor, and of its independence on the specific form of the matter Lagrangian. This result also proves the uniqueness of the predictions of the cosmological models in modified gravity theories, and eliminates the degeneracy specific to the previous approaches. As a particular example of the application of our results we will concentrate on the $f(R,T)$ gravity theory, in which the tensor $\mathbb{T}_{\mu \nu}$ plays an important role, and we will investigate a simple cosmological model with the main goal of giving a first estimation of the implications of the new approach.

The present paper is organized as follows. The general thermodynamic formalism used for the calculation of the second variation of the matter Lagrangian is discussed in Section~\ref{sect2}. The general expression for the second variation of the matter Lagrangian, and of the variation of the energy-momentum tensor is presented in Section~\ref{sect3}. Some cosmological applications of the obtained results are presented in Section~\ref{sect4}. We then briefly review the basics of the $f(R,T)$ gravity theory in Section~\ref{sect1} and outline its cosmological implications for a simple choice $f(R,T)=\alpha |T|^n$. Finally, we discuss and conclude our results in Section~\ref{sect5}.

\section{Thermodynamics and geometry}\label{sect2}

In order to obtain the second variation of the baryonic matter Lagrangian, it is necessary to review the derivation of its first variation using thermodynamics considerations.
 The first law of the thermodynamic is given by
\be
dU=TdS-PdV+\mu dN,\label{first}
\ee
where $U$ is the total energy, $\mu $ is the chemical potential, related to the change in the number of particles in the system, $N$ is the particle number and $V$ is the volume enclosing the fluid.  An important thermodynamic relation is the Gibbs-Duhem equation,
\be
U=TS-PV+\mu N,\label{gibbs1}
\ee
which follows from the extensivity of the energy, $U(\lambda X)=\lambda U(X)$, where $\lambda$ is a constant, and from Euler's theorem of the homogeneous functions.

Let us define the particle number density $n=N/V$ and entropy per particle $s=S/N$.
The first law of thermodynamics \eqref{first} and the Gibbs-Duhem relation \eqref{gibbs1} can be simplified to \cite{brown}
\begin{align}
	d\rho&=Tnds+\mu^\prime dn,\\
	\rho&=\mu^\prime n-P,
\end{align}
where $\mu^\prime=\mu+Ts$ and we have defined the energy density as $\rho=U/V$. Also, by taking the differential of the Gibbs-Duhem relation \eqref{gibbs1} we obtain $$dU=TdS+SdT-PdV-VdP+Nd\mu+\mu dN,$$ and using
the first law of thermodynamics \eqref{first}, one can obtain
\begin{align}
	dP=sndT+nd\mu =nd\mu^\prime-nTds,
\end{align}
implying that $\rho=\rho(s,n)$ and $P=P(\mu^\prime,s)$.

Now, we define the particle number flux
\be
J^\mu=\sqrt{-g}nu^\mu,
\ee
and the Taub current \cite{brown}
\begin{align}
	V_\mu=\mu^\prime u_\mu,
\end{align}
where $u^\mu$ is the fluid 4-velocity, and $n$, the particle number density, can be obtained according to the relation,
\be
n=\sqrt{\frac{g_{\mu \nu }J^{\mu }J^{\nu }}{g}}.
\ee.

From the definition of $\mu^\prime$ one can identify $V^\mu$ with the 4-momentum per particle of a small amount of fluid to be injected in a large sample of fluid without changing the total fluid volume or the entropy per particle \cite{brown}.

With the above definition, one obtains
\begin{align}
	J\equiv\sqrt{-J_\mu J^\mu}=\sqrt{-g}n,\quad J^\mu=Ju^\mu,\label{JJ}\\
	V\equiv\sqrt{-V_\mu V^\mu}=\mu^\prime,\quad V_\mu=Vu_\mu.\label{VV}
\end{align}

In the context of general relativity, it is well-known that there are two equivalent baryonic matter Lagrangians corresponding to
\begin{align}\label{lags}
L_m=-\rho,\qquad L_m=p,
\end{align}

It should be noted that from the definition of the energy-momentum tensor as
\begin{align}\label{EMT}
	T_{\mu\nu}=-\frac{2}{\sqrt{-g}}\frac{\delta(\sqrt{-g}L_m)}{\delta g^{\mu\nu}},
\end{align}
both Lagrangians in Eq.~\eqref{lags} give the same result
\begin{align}\label{EMtensor}
	T_{\mu\nu}=(\rho+P)u_\mu u_\nu+Pg_{\mu\nu}.
\end{align}

It should be noted that the entropy density $s$, the ordinary matter number flux vector
density $J^{\mu }$, and the Taub current $V_\mu$, do not depend on the metric tensor. As a result, their variations with respect to the metric identically vanish \cite{Fel,brown}
 \begin{align}
\frac{ \delta s}{\delta g^{\alpha\beta}}=0,\quad
\frac{\delta J^{\mu }}{\delta g^{\alpha\beta}}=0,\quad
	\frac{\delta V_\mu}{\delta g^{\alpha\beta}}=0.
\end{align}

It is worth mentioning that the derivation of the ordinary matter energy-momentum tensor from the variational principle is implemented with the help of several  constraints \cite{brown}. The famous constraints for a perfect fluid are the conservation of the matter density flux $\nabla_\mu J^\mu=0$, and the conservation of the entropy per particle flux $\nabla_\mu(sJ^\mu)=0$, which are added to the perfect fluid action using Lagrange multipliers. The variations with respect of the Lagrange multipliers then ensure the conservation of the particle number flux, and of the entropy per particle, which finally results in the conservation of the energy-momentum tensor \cite{brown,schutz1970}. In this paper, since we are going to apply the procedure to the cases with non-conserved matter energy-momentum tensor, {\it we will not add these constraints to the perfect fluid action}. As a result, the only constraint we have is $u_\mu u^\mu=-1$, which can be added to the action by using the Lagrange multiplier, or just imposing it during the calculations.

By taking the variation of the particle
number $n$, with the use of the assumptions previously introduced, we find,
\begin{align}\label{14}
\delta n&=\frac{n}{2}\left( -g\right) u^{\mu }u^{\nu }\left( \frac{\delta
g_{\mu \nu }}{g}-\frac{g_{\mu \nu }}{g^{2}}\delta g\right) \notag \\
&=\frac{n}{2}\left( u_{\mu }u_{\nu }+g_{\mu \nu }\right) \delta g^{\mu \nu
}.
\end{align}

In order to obtain the variation of the energy-momentum tensor, we need to find the variations of the energy density and pressure with respect to the metric, namely, $\delta\rho/\delta g^{\mu\nu}$ and $\delta P/\delta g^{\mu\nu}$, respectively.
In the case of isentropic processes, we have
\begin{align}
	\delta\rho&=\frac{\rho+P}{n}\delta n,\\
	\delta P&=n\,\delta\mu^\prime.
\end{align}

Let the equation of state for matter be given as $\rho =\rho \left(
n,s\right) $. Then, since $\delta s=0$, from the thermodynamic relation $%
\left( \partial \rho /\partial n\right) _{s}=w=\left( \rho +P\right) /n$, we
obtain $\delta \rho =w\delta n$.

The variation of $n$ is given by Eq~(\ref{14}), while the variation of  $\mu^\prime$ from equation \eqref{VV} can be obtained as,
%\begin{align}
%	\delta n= \frac{1}{2}n(g_{\mu\nu}+u_\mu u_\nu)\delta g^{\mu\nu},
%\end{align}
%and
\begin{align}\label{17}
	\delta\mu^\prime=\delta V=-\frac{V_{\mu}V_\nu}{2V}\delta g^{\mu\nu}=-\frac12\mu^\prime u_\mu u_\nu \delta g^{\mu\nu}.
\end{align}

These relations give the thermodynamic variations of the energy density and pressure with respect to the metric as,
\begin{align}
	\frac{\delta\rho}{\delta g^{\mu\nu}}&=\frac12(\rho+P)(g_{\mu\nu}+u_\mu u_\nu),\label{18-1}\\
		\frac{\delta P}{\delta g^{\mu\nu}}&=-\frac12(\rho+P)u_\mu u_\nu.\label{18-2}
\end{align}

Eqs.~(\ref{18-1}) and (\ref{18-2}) can be obtained in a direct way by starting from the definition of the matter energy-momentum tensor, as given by Eq.~(\ref{EMT}).
If the matter Lagrangian does not depend on the derivatives of the metric tensor, from Eq.~(\ref{EMT}) we obtain
\be\label{emt}
T_{\mu \nu}=L_mg_{\mu \nu}-2\frac{\delta L_m}{\delta g^{\mu \nu}},
\ee
giving
\be
\frac{\delta L_m}{\delta g^{\mu \nu}}=\frac{1}{2}L_mg_{\mu \nu}-\frac{1}{2}T_{\mu \nu}.
\ee

If we take now $L_m=-\rho$, from the above equation we find
\be
\frac{\delta (-\rho)}{\delta g^{\mu \nu}}=-\frac{1}{2}\rho g_{\mu \nu}-\frac{1}{2}T_{\mu \nu}=-\frac{1}{2}(\rho+P)\left(g_{\mu\nu}+u_\mu u_\nu\right),
\ee
where we have used the expression (\ref{EMtensor}) for the energy-momentum tensor. For $L_m=P$, we obtain
\begin{align}
\frac{\delta P}{\delta g^{\mu \nu}}=\frac{1}{2}Pg_{\mu \nu}-\frac{1}{2}T_{\mu \nu}=-\frac{1}{2}(\rho+P)u_{\mu}u_\nu.
\end{align}
Hence, we have recovered the expressions of the variations with respect to the metric of the energy and pressure variations, previously obtained from first principle thermodynamic considerations.

\section{The first variation of the matter energy-momentum tensor}\label{sect3}

Now, we have all the necessary tools for computing the second variation of the energy density and of the pressure of a perfect fluid.
First, let us note that,
\begin{align}
\delta g_{\mu\nu}=-g_{\mu\alpha}g_{\nu\beta}\delta g^{\alpha\beta}.
\end{align}
In order to obtain the variation of the 4-velocity with respect to the metric,  from the definition $u^\mu=dx^\mu/d\tau$, where,
\begin{align}
	d\tau^2=-g_{\mu\nu}dx^\mu dx^\nu,
\end{align}
one obtains
\begin{align}
	\delta(d\tau^2)=-\delta g_{\mu\nu}dx^\mu dx^\nu.
\end{align}
Then, we find,
\begin{align}
	\delta(d\tau)=\frac{1}{2d\tau}\delta(d\tau^2)=-\frac12\delta g_{\mu\nu}u^\mu dx^\nu.
\end{align}

As a result, the variation of the 4-velocity is obtained as,
\begin{align}
	\delta u^\mu&=\delta\left(\frac{dx^\mu}{d\tau}\right)=-\frac{dx^\mu}{d\tau^2}\delta(d\tau)\nonumber\\&=\frac12u^\mu u^\alpha u^\beta \delta g_{\alpha\beta}=-\frac12 u^\mu u_\alpha u_\beta \delta g^{\alpha\beta}.
\end{align}
Also, one finds,
\begin{align}
	\delta u_\mu=\delta(g_{\mu\nu}u^\nu)=-\frac12(g_{\mu\alpha}u_\beta+g_{\mu\beta}u_\alpha+u_\mu u_\alpha u_\beta)\delta g^{\alpha\beta}.
\end{align}

With these results, it immediately follows that,
\begin{widetext}
\be
	\frac{\delta^2P}{\delta g^{\alpha\beta}\delta g^{\mu\nu}}\equiv \frac{\delta}{\delta g^{\alpha\beta}}\left(\frac{\delta p}{\delta g^{\mu\nu}}\right)
=\frac14(\rho+P)\left(g_{\mu\beta}u_\alpha u_\nu+g_{\mu\alpha}u_\beta u_\nu+g_{\nu\beta}u_\alpha u_\mu+g_{\nu\alpha}u_\beta u_\mu-g_{\alpha\beta}u_\mu u_\nu+2 u_\mu u_\nu u_\alpha u_\beta\right),
\ee
\end{widetext}
and
%\begin{widetext}
\begin{align}
\frac{\delta^2(-\rho)}{\delta g^{\alpha\beta}\delta g^{\mu\nu}}&=\frac{\delta^2P}{\delta g^{\alpha\beta}\delta g^{\mu\nu}}\nonumber\\
&-\frac14(\rho+P)(g_{\alpha\beta}g_{\mu\nu}-g_{\mu\alpha}g_{\nu\beta}-g_{\mu\beta}g_{\nu\alpha}),
\end{align}
%\end{widetext}
respectively.

After a little algebra, and by assuming that the matter Lagrangian does not depend on the derivatives of the metric tensor,  one can obtain from its definition (\ref{EMT}) the variation of the energy-momentum tensor as,
%\begin{widetext}
\begin{align}
	\frac{\delta T_{\mu\nu}}{\delta g^{\alpha\beta}}&=\frac12L_m(g_{\alpha\beta}g_{\mu\nu}-g_{\mu\alpha}g_{\nu\beta}-g_{\mu\beta}g_{\nu\alpha})\nonumber\\
&-\frac12T_{\alpha\beta}g_{\mu\nu}-2\frac{\delta^2L_m}{\delta g^{\alpha\beta}\delta g^{\mu\nu}}.
\end{align}
%\end{widetext}

Therefore, after substituting the expressions of the second variations of the matter Lagrangians, we find {\it the important result that for both baryonic matter Lagrangians in} Eq.~\eqref{lags}, we obtain,
%\begin{widetext}
\begin{align}\label{res1}
		\frac{\delta T_{\mu\nu}}{\delta g^{\alpha\beta}}&=\frac12P (g_{\alpha\beta}g_{\mu\nu}-g_{\mu\alpha}g_{\nu\beta}-g_{\mu\beta}g_{\nu\alpha})\nonumber\\
&-\frac12T_{\alpha\beta}g_{\mu\nu}-2\frac{\delta^2P}{\delta g^{\alpha\beta}\delta g^{\mu\nu}},
\end{align}
%\end{widetext}
implying that {\it the expression of} $\delta T_{\mu\nu}/\delta g^{\alpha\beta}$ {\it is independent on the choice of the matter Lagrangian}.
This is not the case for the approximate result obtained by neglecting the second variation of the matter Lagrangian with respect to the metric,
\begin{align}
	\frac{\delta T_{\mu\nu}}{\delta g^{\alpha\beta}}\approx \frac12L_m(g_{\alpha\beta}g_{\mu\nu}-g_{\mu\alpha}g_{\nu\beta}-g_{\mu\beta}g_{\nu\alpha})-\frac12T_{\alpha\beta}g_{\mu\nu},
\end{align}
which obviously depends on the choice of Lagrangian density.

{\it It should be noted at this moment that the energy-momentum tensor, and its variation, should be independent to the choice of the baryonic matter Lagrangian}, as we have summarized in the previous Section, on thermodynamics grounds.

Eq.~\eqref{res1} can also be written in the form,
%\begin{widetext}
\bea\label{rescor}
\frac{\delta T_{\mu\nu}}{\delta g^{\alpha\beta}}&=&\frac12P(g_{\nu\beta}g_{\alpha\mu}+g_{\nu\alpha}g_{\beta\mu})-(\rho+P)u_\mu u_\nu u_\alpha u_\beta\nonumber\\
&&-\frac12\Bigg(T_{\alpha\nu}g_{\mu\beta}+T_{\beta\nu}g_{\mu\alpha}+T_{\alpha\mu}g_{\nu\beta}+T_{\beta\mu}g_{\nu\alpha}\nonumber\\
&&-T_{\mu\nu}g_{\alpha\beta}+T_{\alpha\beta}g_{\mu\nu}\Bigg).
\eea
%\end{widetext}

Also, by defining a modified energy-momentum tensor,
\begin{align}\label{mod}
\bar{T}_{\mu\nu}=(\rho+P)u_\mu u_\nu+\frac12Pg_{\mu\nu},
\end{align}
one can write the first variation of the energy-momentum tensor as,
\begin{widetext}
\begin{align}
	\frac{\delta T_{\mu\nu}}{\delta g^{\alpha\beta}}=-\frac12\left(\bar{T}_{\beta\nu}g_{\mu\alpha}+\bar{T}_{\alpha\nu}g_{\mu\beta}+\bar{T}_{\alpha\mu}g_{\nu\beta}+\bar{T}_{\beta\mu}g_{\nu\alpha}-\bar{T}_{\mu\nu}g_{\alpha\beta}+\bar{T}_{\alpha\beta}g_{\mu\nu}\right)-(\rho+P)u_\mu u_\nu u_\alpha u_\beta.
\end{align}
\end{widetext}

‌‌‌In the well-known $f(R,T)$ gravity theories \cite{fRT}, on encounters with the expression $g^{\mu\nu}\delta T_{\mu\nu}/\delta g^{\alpha\beta}$, which enters into the modified field equations. With the result given by Eq.~\eqref{rescor}, we define,
\begin{align}\label{36}
	\mathbb{T}_{\alpha\beta}\equiv g^{\mu\nu}\frac{\delta T_{\mu\nu}}{\delta g^{\alpha\beta}}
	=-3(\rho+P)u_\alpha u_\beta-\frac12(\rho+3P)g_{\alpha\beta}.
\end{align}
Alternatively, we also have,
\begin{align}
	\frac{\delta T}{\delta g^{\alpha\beta}}=T_{\alpha\beta}+\mathbb{T}_{\alpha\beta}.
\end{align}

In the comoving frame, one can then obtain,
\begin{align}\label{termcor}
	\mathbb{T}^\mu_\nu=\frac12\textmd{diag}\left(5\rho+3P, -\rho-3P,-\rho-3P,-\rho-3P\right).
\end{align}

Taking the trace of the above expression, one finds,
\begin{align}
	\mathbb{T}\equiv g^{\mu\nu} \mathbb{T}_{\mu\nu}= (\rho-3P).
\end{align}

The approximate results, obtained by neglecting the second variation of the matter Lagrangian, is
\begin{align}
	\mathbb{T}_{\alpha\beta} \approx (L_m-2P)g_{\alpha\beta}-2(\rho+P)u_\alpha u_\beta,
\end{align}
which could be simplified as
\begin{align}
	\mathbb{T}^\mu_\nu \approx \textmd{diag}\left(\rho, -\rho-2P,-\rho-2P,-\rho-2P\right),
\end{align}
for $L_m=-\rho$, and
\begin{align}
	\mathbb{T}^\mu_\nu \approx \textmd{diag}\left(2\rho+P, -P,-P,-3P\right),
\end{align}
for $L_m=P$.

For the approximate result with $L_m=-\rho$ we obtain $\mathbb{T}\approx -2(\rho+3P)$, while for $L_m=P$ we obtain $\mathbb{T}\approx 2(\rho-P)$.  In Fig.~\ref{fig1} we have plotted the new exact result, together with the result from previous considerations.

\subsection{Cosmological implications}\label{sect4}

In order to determine the effect of the new term in the variation of the energy-momentum tensor, let us find its behavior for a conserved matter source in a flat FLRW Universe, with the line element
\begin{align}
	ds^2=-dt^2+a^2(t)\left(dx^2+dy^2+dz^2\right),
\end{align}
where $a$ is the scale factor.

In this case, one has for the baryonic matter density $\rho_m$, assumed to be in the form of dust, the expression
\begin{align}
	\rho_m=\frac{\Omega_{m0}}{a^3},
\end{align}
where $\Omega_{0m}$ is the present time density abundance. For the variation of the density of the radiation we have
\be
\rho _r=\frac{\Omega _{r0}}{a^4}.
\ee

Assume that the Universe is filled with dust and radiation, with
\begin{align}
	\rho=\rho_m+\rho_r=\frac{\Omega_{m0}}{a^3}+\frac{\Omega _{r0}}{a^4},\quad P=\frac13\rho_r.
\end{align}

In this case, one obtains
\begin{align}
	\mathbb{T}=\Omega_{m0}(1+z)^3,
\end{align}
where we have introduced the redshift $z$, defined as
\begin{align}
	1+z=\frac1a,
\end{align}
and $\Omega_{m,0}$ and $\Omega_{r,0}$ are the current values of the dust and radiation abundances, $\Omega_{m0}=0.305$, and $\Omega_{r0}=5.3\times 10^{-5}$, respectively \cite{hubble}.

\begin{figure}[t]
	\includegraphics[scale=0.9]{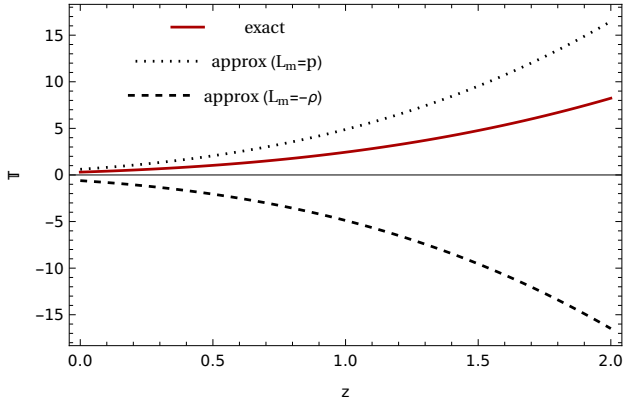}
	\caption{The behavior of the extra term $\mathbb{T}$ as a function of the redshift $z$ for the new exact expression (solid curve), and for the previously considered approximate relations, for $L_m=-\rho$ (dashed curve), and $L_m=P$ (dotted curve), respectively. \label{fig1}}
\end{figure}

In Fig.~\ref{fig1} we have depicted the evolution of the new term $\mathbb{T}$ as a function of the redshift. As a result, we expect that the new term changes the behavior of the cosmological models in theories in which the first order variation of the energy-momentum tensor with respect to the metric is present in the gravitational field equations. There are major differences as compared with the approximate relation for $L_m=-\rho$, but the two relations coincide for $L_m=P$.

\section{$f(R,T)$ gravity}\label{sect1}

Now let us consider a typical gravitational theory in which the above results can have an important influence. Consider the action \cite{fRT},
\begin{align}\label{act11}
	S=\int d^4x\sqrt{-g}(\kappa^2R+f(R,T)+L_m),
\end{align}
where $f(R,T)$ is an arbitrary function of the Ricci scalar $R$, and of the trace of the energy-momentum tensor $T$. We suppose that the Universe is filled with a perfect fluid with the matter energy-momentum  having the form \eqref{EMtensor}. The field equations can be obtained as
\begin{align}\label{fieldeq}
	\kappa^2G_{\mu\nu}&-\frac12 f g_{\mu\nu}+f_R R_{\mu\nu}+(g_{\mu\nu}\Box-\nabla_\mu\nabla_\nu)f_R\nonumber\\&=\frac12T_{\mu\nu}-f_TT_{\mu\nu}-f_T\mathbb{T}_{\mu\nu},
\end{align}
where the last term is computed as in Eq.~\eqref{36}. It should be noted that using the exact result Eq.~\eqref{36}, {\it the choice of the matter Lagrangian is irrelevant, both cases with $L_m=-\rho$ and $L_m=P$ giving the same field equations}.

\begin{figure*}[htbp]
	\includegraphics[scale=0.5]{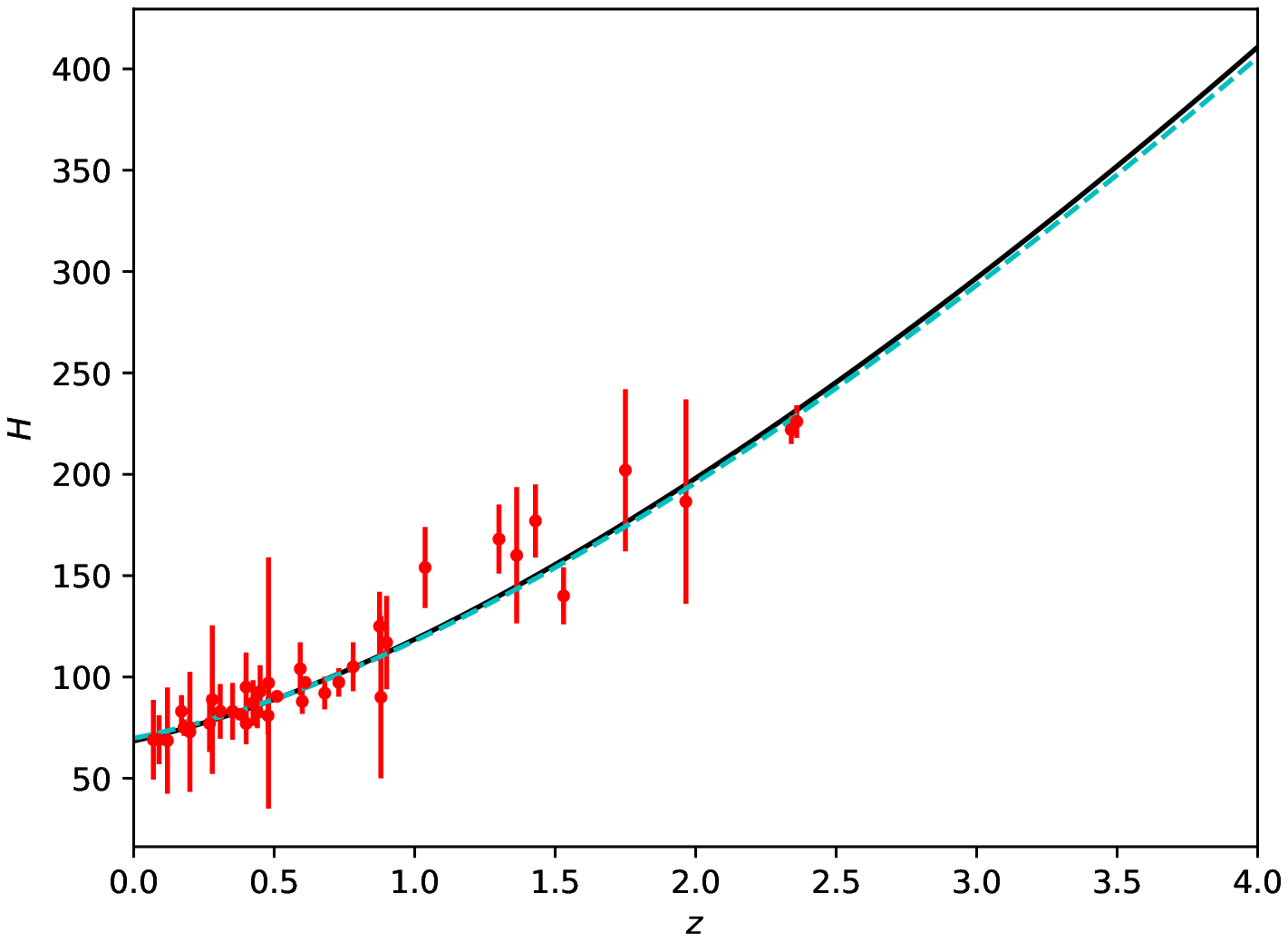}\includegraphics[scale=0.5]{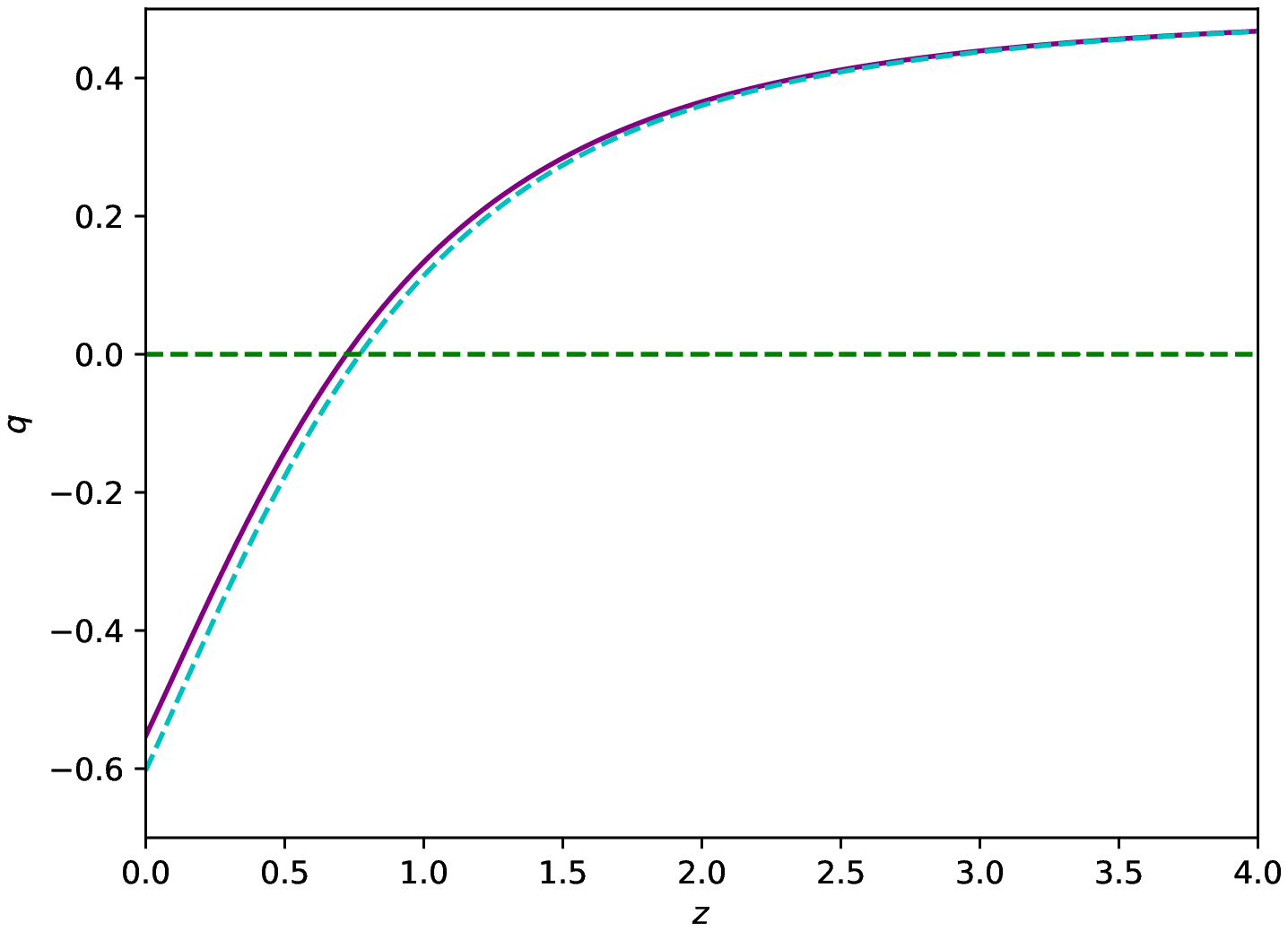}
	\caption{The behavior of the Hubble parameter $H$ and of the deceleration parameter $q$ as a function of the redshift for the best fit values of the parameters as given by Eqs.~(\ref{best}). The dashed line represents the $\Lambda$CDM model.\label{fig2}}
\end{figure*}

\begin{figure}[htbp]
	\includegraphics[scale=0.5]{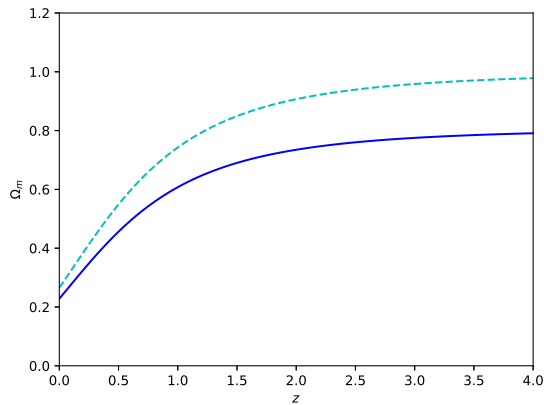}
	\caption{The behavior of the matter density parameter $\Omega_{m}$ as a function of redshift for the best fit values of the parameters as given by  Eq.~(\ref{best}). The dashed line represents the $\Lambda$CDM model.\label{fig3}}
\end{figure}
\begin{figure}[htbp]
	\includegraphics[scale=0.5]{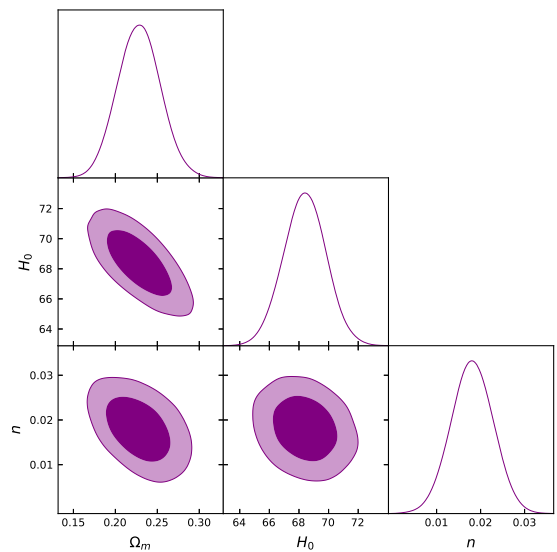}
	\caption{The corner plot for the values of the parameters $H_0$, $\Omega_{m0}$ and $n$ with their $1\sigma$ and $2\sigma$ confidence levels. \label{trig}}
\end{figure}

With the use of the mathematical identity $$\left(\square\nabla_\nu-\nabla_\nu\square\right)f_R=R_{\mu\nu}\nabla^\mu f_R,$$ after taking the divergence of Eq.~\eqref{fieldeq} we obtain the conservation equation in the $f\left(R,T\right)$ gravity theory in the form
\begin{align}\label{eq:conserv-general}
	\left(\frac12-f_T\right)\nabla^\mu T_{\mu\nu}&=\left(T_{\mu\nu}+\mathbb{T}_{\mu\nu}\right)\nabla^\mu f_T \nonumber\\&
	+f_T\left(\nabla^\mu\mathbb{T}_{\mu\nu}+\frac12\nabla_\nu T\right).
\end{align}

As one can see from the field equations (\ref{fieldeq}), the dynamical behavior in $f(R,T)$ gravity essentially depends on the tensor $\mathbb{T}_{\mu \nu}$.
In this paper, we will consider a simple case that indicates the importance of the new term. Let us assume that $f(R,T)=\alpha |T|^n$, and $P=0$. In this case, the field equations reduce to
\begin{align}
	\kappa^2 G_{\mu\nu}=\frac12 T_{\mu\nu}+\frac12\alpha |T|^n g_{\mu\nu}-n\alpha\epsilon|T|^{n-1}(T_{\mu\nu}+\mathbb{T}_{\mu\nu}),
\end{align}
where $\epsilon=sign(T)$. Here we have $T=-\rho$ and then $\epsilon=-1$.
The Friedmann and Raychaudhuri equations are then
\begin{align}
h^2&=\bar\rho_m-\beta(3n+1)\bar\rho_m^n,\label{frid}\\
h^\prime&=-\frac32\left(\bar\rho_m-4\beta n \bar\rho_m^n\right),
\end{align}
where we have used the following set of dimensionless variables,
\begin{align}
		\tau&=H_0t,\quad H=H_0 h,\nonumber\\
		 \bar\rho&=\frac{\rho}{6\kappa^2 H_0^2},\quad \beta=(6\kappa^2H_0^2)^{n-1}\alpha,
\end{align}
and we have denoted by $H_0$ the current value of the Hubble parameter, and by a prime the derivative with respect to $\tau$. As an indicator of the decelerating/accelerating evolution we introduce the deceleration parameter, defined as
\be
q=\frac{d}{d\tau}\frac{1}{h}-1.
\ee

Note that from the normalized Friedmann equation \eqref{frid}, and by taking into account that at the present time we have $h(present)=1$, we can obtain the coupling $\beta$ as
\begin{align}\label{beta}
	\beta=-\frac{1-\Omega_{m0}}{(1+3n)\Omega_{m0}^n}.
\end{align}

 In order to find the best fit value of the parameter $n$, $H_0$ and $\Omega_{m0}$, we use the Likelihood analysis using the observational data on the Hubble parameter in the redshift range $z\in(0.07,2.36)$ \cite{hubble}.  In the case of independent data points, the likelihood function can be defined as
\begin{align}
	L=L_0e^{-\chi^2/2},
\end{align}
where $L_0$ is the normalization constant and the quantity $\chi^2$ is defined as
\begin{align}
	\chi^2=\sum_i\left(\frac{O_i-T_i}{\sigma_i}\right)^2.
\end{align}
Here $i$ counts the data points, $O_i$ are the observational value, $T_i$ are the theoretical values, and $\sigma_i$ are the errors associated with the $i$th data obtained from observations.

By maximizing the likelihood function, the best fit values of the parameters $n$, $\Omega_{m0}$ and $H_0$ at $1\sigma$ confidence level, can be obtained as
\begin{align}\label{best}
	\Omega_{m0}&=0.228^{+0.025}_{-0.025},\nonumber\\
	H_0&=68.396^{+1.421}_{-1.437},\nonumber\\
	n&=0.018^{+0.005}_{-0.005}.
\end{align}
Also, with the use of equation \eqref{beta} we obtain
\begin{align}
	\beta=-0.756^{+0.031}_{-0.030}.
\end{align}

The redshift evolution of the Hubble function, of the deceleration parameter $q$, and of the matter density parameter $\Omega_m=\bar\rho_m/h^2$ are represented, for this model, in Figs.~\ref{fig2} and \ref{fig3}, respectively. Also, the corner plot for the values of the parameters $H_0$, $\Omega_{m0}$ and $n$ with their $1\sigma$ and $2\sigma$ confidence levels is shown in Fig.~\ref{trig}.

\section{Discussions and final remarks}\label{sect5}

In the present paper we have obtained the complete expression of the first variation of the matter energy-momentum tensor with respect to the metric $g^{\mu \nu}$, and of its associated tensor $\mathbb{T}_{\mu \nu}$. The full estimation of this term requires the calculation of the second variations of the matter Lagrangian with respect to the metric, a term which was generally ignored in the previous investigations of this problem. The expression of $\delta ^2L_m/\delta g^{\mu \nu}\delta g^{\alpha \beta}$ can be calculated straightforwardly from the first variation $\delta L_m/\delta g^{\mu \nu}$, which can be obtained for the two possible choices of the matter Lagrangian either from thermodynamic considerations, or in a direct way by using the definition of the energy-momentum tensor. The main result of this work is that the first variation of the matter energy-momentum tensor, given by Eq.~(\ref{res1}), is independent of the choice of the matter Lagrangian; both possible choices lead to the same expression (\ref{res1}), depending only on the thermodynamic pressure, and its second variation. The variation of the energy-momentum tensor can also be expressed in terms of the pressure, and the energy-momentum tensor itself, or in a compact form in terms of a generalized energy-momentum tensor, formally defined in Eq.~(\ref{mod}).

The new form of the variation of the matter energy-momentum tensor may have some important implications on modified gravity theories with geometry-matter coupling. As an important example we have considered the particular case of the $f(R,T)$ gravity theory. We have investigated the cosmological implications of a particular representation of the $f(R,T)$ gravity, with action given by Eq.~(\ref{act11}), in which the standard Hilbert-Einstein Lagrangian is replaced by a general term $f(R,T)$. As a simple case we have taken $f(R,T)=\alpha |T|^n$. The generalized Friedmann equations take a simple form, and they allow a complete analysis of the cosmological features of this simple model, and a full fitting of the observational cosmological data, which permits the determination of the optimal values of the free parameters. The model gives an excellent description of the observational data for the Hubble function, up to a redshift of $z\approx 4$ (see \cite{ht1,ht2,ht3,ht4,ht5,ht6} for discussions on the Hubble tension in modified gravities.). In this redshift range the model basically coincides with the $\Lambda$CDM model. The transition from acceleration to deceleration takes place a redshift that again coincides with the $\Lambda$CDM value. Moreover, the deceleration parameter $q$ basically coincides with the $\Lambda$CDM prediction. However, significant differences in the behavior of the matter density do appear at higher redshifts.

It is worth discussing the relation of the present theory with a general scalar-tensor theory of gravity to answer the question that whether we can use a scalar field to describe a general perfect fluid or not. Suppose that we have a scalar theory with an action
\begin{align}
	S=\int d^4x\sqrt{-g}\left(-\frac12g^{\mu\nu}\partial_\mu\phi\partial_\nu\phi+V(\phi)\right),
\end{align}
where $V$ is an arbitrary algebraic function of the scalar field $\phi$. If we consider this action as a matter source for the theory, one can obtain the energy-momentum tensor as
\begin{align}
	T_{\mu\nu}=-\frac12\partial_\alpha\phi\partial^\alpha\phi g_{\mu\nu}+Vg_{\mu\nu}+\partial_\mu\phi\partial_\nu\phi,
\end{align}
which can be written in the form of a perfect fluid with energy density and pressure of the form
\begin{align}
	\rho&=-\frac12\partial_\alpha\phi\partial^\alpha\phi-V,\nonumber\\
	p&=-\frac12\partial_\alpha\phi\partial^\alpha\phi+V
\end{align}
and
\begin{align}
	u_\mu=\frac{\partial_\mu\phi}{\sqrt{-\partial_\alpha\phi\partial^\alpha\phi}}.
\end{align}
At first sight, one can see that the above matter source can be seen as a perfect fluid with $L_m=p$. This may arise the question that whether we can use a scalar field to obtain the relations and results  of the present paper, instead of performing the full thermodynamics calculations. In order to give a definite answer to this question we would like to point out that there are some important differences, from the thermodynamical point of view, between the perfect fluid matter sources, and the scalar field configurations considered  above.
First of all, unlike in the thermodynamics case, one can easily verify that the Lagrangian $L_m=-\rho$ for the scalar field does not produce the correct energy-momentum tensor \cite{farao}. This is because the scalar field is a special case of a perfect fluid, with definite equation of state, say $p=\rho+2V$. It should be noted at this moment that in order to relate the scalar field and a perfect fluid, the scalar field Lagrangian should be shift-symmetric \cite{TM2}, a condition which implies $V=0$. Thus, the scalar field  behaves as stiff matter, with $p=\rho$. One should also emphasize that in order to be able to formulate general statements about perfect fluids, one should consider the general case of a perfect fluid, without imposing any specific equation of state. There is a concrete example about this aspect, which may clarify the issue. Suppose we consider a perfect fluid for which we have proven that both Lagrangians $L_m=-\rho$ and $L_m=p$ give the same energy-momentum tensor. Now, impose an equation of state of the form $p=\alpha \rho^n$ explicitly in the action. As a result, the Lagrangian $L_m=p$ becomes $L_m=\alpha\rho^n$, which obviously does not give us the right energy-momentum tensor as obtained from  $L_m=-\rho$. The equation of state for the matter source must be imposed only after obtaining the equations of motion, since otherwise we obtain a wrong expression for the matter energy-momentum tensor. In fact this is what happens when we try to consider the scalar field as a perfect fluid source in the action. A related argument about the incompatibility of the scalar description for a perfect fluid from the EFT perspective has been discussed in \cite{thermo}.
The second reason why the scalar field can not be used as a perfect fluid source in the action is that a perfect fluid has two independent thermodynamic variables, namely, the specific entropy $s$, and the number density $n$, respectively \cite{thermo}. On the other hand, the scalar field is described by one independent variable only, namely, $\phi$. Hence, we can not obtain all the properties of the perfect fluids from the scalar field Lagrangian. In fact, in \cite{TM2}, the authors have discussed that the shift-invariant scalar fields can describe accurately the potential flow of an isentropic perfect fluid, but, in general, the identification is possible only for a finite period of time.

The calculations and the results of the present paper give the expression of the variation of the energy-momentum tensor in the cases where we have particle creation, a situation which is typical for theories with non-minimal couplings between matter and geometry. This means that the present approach is valid for theories with non-conserved energy-momentum tensor $\nabla_\mu T^{\mu\nu}\neq0$. In the case where we want to investigate particle creation in a theory with conserved energy-momentum tensor, we have to define some extra thermodynamic quantities, like, for example, a creation pressure (in the case of symmetric space-time), in order to describe the properties of particle creation with conserved energy-momentum tensor. For a discussion and study of such cases see \cite{lima}, for example. In this kind of physical situations, the form of the energy-momentum tensor differs from that of perfect fluid; the extra terms appearing in $T_{\mu \nu}$ play the role of the right hand side of the non-conservation equation of the energy-momentum tensor in modified gravity theories.

The search for the “true” physical quantities from which the matter energy-momentum tensor can be obtained ($-\rho$ or $P$) in a variational formulation is still going on. Interestingly enough, the  two possible matter Lagrangians are not equivalent in any sense (physical or mathematical), but their functional variation coincides, leading to the same energy-momentum tensor. 

A model in which particles are described as localized concentrations of energy, with fixed rest mass and structure, which are not significantly affected by their self-induced gravitational field, was considered in \cite{N18a}.  The volume average of the on-shell matter Lagrangian $L_m$ describing such particles turns out to be equal to the volume average of the trace $T$ of the energy-momentum tensor, independently of the particle's structure and constitution. These results are relevant for theories in which the matter Lagrangian appears explicitly in the field equations, such as the $f\left(R,L_m\right)$ and $f(R,T)$ gravity theories. They also indicate that $f\left(R,L_m\right)$ theories could be interpreted as a subclass of the $f(R,T)$ gravity. In \cite{N18b} it was shown that the on-shell Lagrangian of a perfect fluid depends on the microscopic properties of the fluid. If the fluid consists of localized concentrations of energy with fixed rest mass and structure (solitons), then the average on-shell Lagrangian of a perfect fluid is given by $L_m=T$. These results could lead to observable deviations from a nearly perfect cosmic microwave background black body spectrum.

In \cite{N20a} it was proven that generally the density-pressure degeneracies are broken for nonminimal coupled fluids. Thus, in these cases, the determination of the appropriate on-shell Lagrangian is essential for the description of the overall dynamics. Models with the same on-shell Lagrangian may have different proper energy densities. This result can be used to map dark matter models into unified dark energy models, with dark matter and dark energy described by the same perfect fluid.

In \cite{N21a} it was shown that the matter Lagrangian of an ideal fluid is equal (up to a sign depending on its definition, and on the chosen signature of the metric) to the total energy density of the fluid, that is, the rest energy density plus the internal energy density.
By considering the dynamics of particles and fluids in the context of theories of gravity nonminimally coupled to matter, in \cite{N22a} it was shown that the two most common choices for the on-shell Lagrangian of a perfect fluid ($L=P$ and $L=-\rho$, do not apply to an ideal gas, except in the case of dust, satisfying the condition $T =-\rho$.

 For a wide class of modified fluid Lagrangians, the
variational principle is unable to give the complete set of field equations, and usually  one additional equation is required
for completeness. In \cite{N23a} it was shown that by adopting the local energy conservation equation, one can obtain a modified fluid source having a good
thermodynamic interpretation. Such a modified Lagrangian gives the Friedmann equations as $3H^2=-kc^2L_F$, and $3\ddot{a}/a=-kc^2\left[L_F-(3n/2)\left(dL_f/dn\right)\right]$, where $L_f$ is the modified fluid Lagrangian. If $L_f = L_f(n)$, then the system of the Friedmann equations is complete, as there are two equations and two variables ${a, n}$. However, if $L_f = L_f(\rho)$, then the cosmological evolution equations are not closed, since no equation determines the evolution of $\rho$. In this case, one supplementary evolution equation is required.  In \cite{N23b} a method to obtain the stress-energy tensor of the perfect fluid was introduced, by adding a suitable term to the Einstein-Hilbert action. The term can be interpreted physically as pressure, and its variation should be taken with respect to the metric.

However, as shown in the present paper, the first variation of the matter energy-momentum tensor is independent on the adopted form of the matter Lagrangian, making the modified gravity theories containing this term unique, and well defined. Hence, the study of the various orders of variations of the matter Lagrangians and of the energy-momentum tensor turns out to be an important field of research, which could lead to a new understanding of the mathematical
formalism,  and of the astrophysical and cosmological implications of the modified gravitational theories, and in particular of the $f(R, T )$ gravity.

\section*{Acknowledgments}
We would like to thank the two anonymous referees for comments and suggestions that helped us to significantly improve our manuscript. The work of TH is supported by a grant of the Romanian Ministry of Education and Research, CNCS-UEFISCDI, project number PN-III-P4-ID-PCE-2020-2255 (PNCDI III).

\end{document}